\documentclass[final,twocolumn]{elsarticle}

\usepackage{lineno,hyperref}
\modulolinenumbers[5]

\journal{ }

\usepackage{graphicx}
\usepackage[figuresright]{rotating}
\usepackage{dblfloatfix}    
\usepackage{mathptmx}       
\usepackage{helvet}         
\usepackage{courier}        
\usepackage{type1cm}        
\usepackage{makeidx}         
\usepackage{multicol}        
\usepackage[bottom]{footmisc}
						
\usepackage{listings}
\usepackage[utf8]{inputenc}
\usepackage{amsmath}
\usepackage[linesnumbered,ruled,vlined]{algorithm2e}

\title{Big Data Model Simulation  on a Graph Database for Surveillance in Wireless Multimedia Sensor Networks}
	
\begin{document}
	
\begin{frontmatter}
		
\author[metu]{Cihan Küçükkeçeci}
\ead{cihan.kucukkececi@ceng.metu.edu.tr}

\author[metu]{Adnan Yazıcı}
\ead{yazici@ceng.metu.edu.tr}

\address[metu]{Middle East Technical University, Department of Computer Engineering, Ankara, Turkey}

\begin{abstract}
Sensors are present in various forms all around the world such as mobile phones, surveillance cameras, smart televisions, intelligent refrigerators and blood pressure monitors. Usually, most of the sensors are a part of some other system with similar
sensors that compose a network. One of such networks is composed of millions of sensors connect to the Internet which is called Internet of things (IoT). With the advances in wireless communication technologies, multimedia sensors and their networks are expected to be major components in IoT. Many studies have already been done on wireless multimedia sensor networks in diverse domains like fire detection, city surveillance, early warning systems, etc. All those applications position sensor nodes and collect their data for a long time period with real-time data flow, which is considered as big data. Big data may be structured or unstructured and needs to be stored for further processing and analyzing. Analyzing multimedia big data is a challenging task requiring a high-level modeling to efficiently extract valuable information/knowledge from  data. In this study, we propose a big database model based on graph database model for handling data generated by wireless multimedia sensor networks. We introduce a simulator to generate synthetic data and store and query big data using graph model as a big database. For this purpose, we evaluate the well-known graph-based NoSQL databases, Neo4j and OrientDB, and a relational database, MySQL. We have run a number of query experiments on our implemented simulator to show that which database system(s) for surveillance in wireless multimedia sensor networks is efficient and scalable.
\end{abstract}


\end{frontmatter}

\section{Introduction}
\label{sec:1}
A wireless multimedia sensor network (WMSN) is a distributed wireless network that consists of a set of multimedia sensor nodes, which are connected to each other or connected to leading gateways. Nowadays, smart devices such as mobile phones, smart televisions, and smart watches are equipped with sensors and network connections. Hence, with the advances in wireless communication technologies, multimedia sensor networks are expected to be one of the major components in the Internet of things (IoT).

A typical application for a WMSN is a surveillance system or a monitoring system. Smart city surveillance cameras with 7/24 recording, or one million sensor nodes reporting meteorological data produce data in various formats as video, audio, and text \cite{perera2014sensing}. All that huge structured or unstructured data is considered as big data, which is defined by a number of Vs; \emph{Volume, Velocity, Variety, Veracity, and Value}. Min et al. \cite{chen2014big} present a comprehensive survey of big data and they identify that “defining the structural model of big data” is a fundamental problem. Fusing and analyzing big data are challenging tasks and there are many research studies that are related to big data from different points of view in recent years. As pointed out by many researchers, relational database management systems (RDBMS) are inadequate for efficiently handling big data; therefore NoSQL database systems are mostly utilized \cite{moniruzzaman2013nosql,vicknair2010comparison,xu2015semantic}. There are four main types of NoSQL databases, which are key-value store (e.g. Amazon’s Simple DB), big table (e.g. Apache Cassandra), document store (e.g. MongoDB) and graph-based model (e.g. Neo4j) \cite{han2011survey}. 

Graph databases consist of nodes and edges (relations between nodes) which store data as properties. Graph databases are very efficient and convenient to handle social networks, fraud detection, graph-based operations, real-time recommendations and hierarchical relations.
While storage of the big data is an important task, processing the streaming data and taking action for mission-critical applications are crucial. Arkady et al. \cite{zaslavsky2013sensing} state that analyzing and extracting the valuable data from dirty raw data is an important research topic. In order to process that kind of data, we need to identify the data flow.

In this paper, we propose to use a graph-based model for handling big data generated from surveillance applications in wireless multimedia sensor networks. For this reason, we propose a graph-based model as a generic model to be used for different surveillance applications. Big sensor data is stored in a graph database for the purpose of advanced analytics, such as data mining, prediction, and statistics. Our graph model represents both the data flow among the nodes and wireless multimedia sensor network topology. The applicability of our solution is illustrated with a prototype implementation including simulation of synthetic data. A case study in the military surveillance domain is simulated and several experiments are done to measure the efficiency of our solution. Simulation results show that our proposed multimedia wireless sensor network model is applicable in large-scale real-life application scenarios.

The contribution of our study is to store the multimedia sensor network data in a well-defined graph-based big database model. The big data stored in an open-source graph database can be used for analyzing, filtering, aggregating and correlating big data. A simulation infrastructure is implemented for simulating multimedia wireless sensor networks to run a number of complex experimental queries. Although there have been some related studies in literature about the surveillance systems in the big data context, to the best of our knowledge, there has not been any applicable graph-based big data model for WMSNs based on a graph database yet.

This article is organized as follows: next two sections provide background information and related work. Then we introduce our real WSN system to give technical details of our deployment. Section \ref{sec:3} is the proposed graph-based big model explanation and Section \ref{sec:4} presents the prototype implementation of our model. Simulation infrastructure of wireless multimedia sensor networks is given in Section \ref{sec:6sim}. Section \ref{sec:5} illustrates a case study in the military surveillance domain and Section \ref{sec:6} presents the experimental results and evaluations. Finally, conclusions are drawn in Section \ref{sec:7}.

\section{Background}
\subsection{Internet of Things}
The Internet of Things (IoT) is used by Kevin Ashton in 1999 \cite{ashton2009internet} and it roughly means that the Things use the Internet instead of Humans. The sensors, RFIDs, and nanotechnology help this mission to be accomplished by taking away the need for human-entered data. 

Pankesh et al.\cite{patel2011towards} propose a domain model for IoT to make a common understanding. To define the model, they reference to the real world applications and summarize under three headings; Intermittent Sensing, Regular Data Collection, and Sense-Compute-Actuate loops.

As parallel to Sense-Compute-Actuate cycle, Sensor-as-a-Service (Senaas) notion is defined by Sarfraz et al. \cite{alam2010senaas}. They virtualized the sensors as services by an abstraction on technical details of sensors. They trigger services with an event in the sensor and compute it to reply an action. Their IoT virtualization framework is validated by a case study.

Atzori et al. \cite{atzori2010internet} prepare a comprehensive survey about IoT. They identify the enabling technologies as sensing, identification and communication systems like RFID, WiFi, and sensors. In addition, middleware applications using Service Oriented Architecture (SOA) are important for data distribution. In the end, they list open issues such as privacy, addressing of things and non-standardized applications for the future.

IoT is fully connected to sensor technology and the researches about sensor networks directly or indirectly improve the IoT. Hong et al. \cite{hong2010snail} propose an approach to IoT using IP-based wireless sensor networks. They realize that IoT probably has the same problems that Internet itself had in the past. So, they identify problems like IPv6 adaptation, mobility, web enablement, global time synchronization, and security. They also share evaluation results of an implementation of their proposed SNAIL (Sensor Networks for an All-IP World) platform.

\subsection{Sensor Networks}
A set of sensors called sensor nodes connected to each other or connected to leading gateways is simply called a sensor network (SN). If sensors have capability of collecting multimedia data and have a communication infrastructure among the sensors, it is called multimedia sensor network (MSN). If sensors are connected to each other using wireless technology then it is a wireless multimedia sensor network (WMSN).  

Akyıldız et al. \cite{akyildiz2007survey} discuss the state of the art of research on WMSNs as well as the challenges. The challenges related to WSN deployment configurations are summarized by Perera et al. \cite{perera2014sensor} in their research. Another survey paper \cite{li2014middleware} enlists the challenging issues to design middleware systems for WSN. Some of the identified challenges are as follows:  data fusion, resource management, scalability and network topology, security, Quality of Service and limited power.

Peng et al. \cite{zhang2013novel} propose a wireless sensor network in which sink node is replaced by a cloud. They called sink point instead of gateway. Their simulation results show that cloud based architecture increases the WSN performance.

Arati et al. focus on the information retrieval from sensor networks and propose a hybrid protocol which is called APTEEN \cite{manjeshwar2002apteen}. They make experiments by executing queries to show that proposed protocol performs better.

From the database point of view, Ramesh et al. \cite{govindan2002sensor} define a database layer on top of sensor network so that a database query is mapped to traversing sensor nodes in the WSN.

\subsection{Big Data}
\begin{figure*}
	\centering
	\includegraphics[scale=0.9]{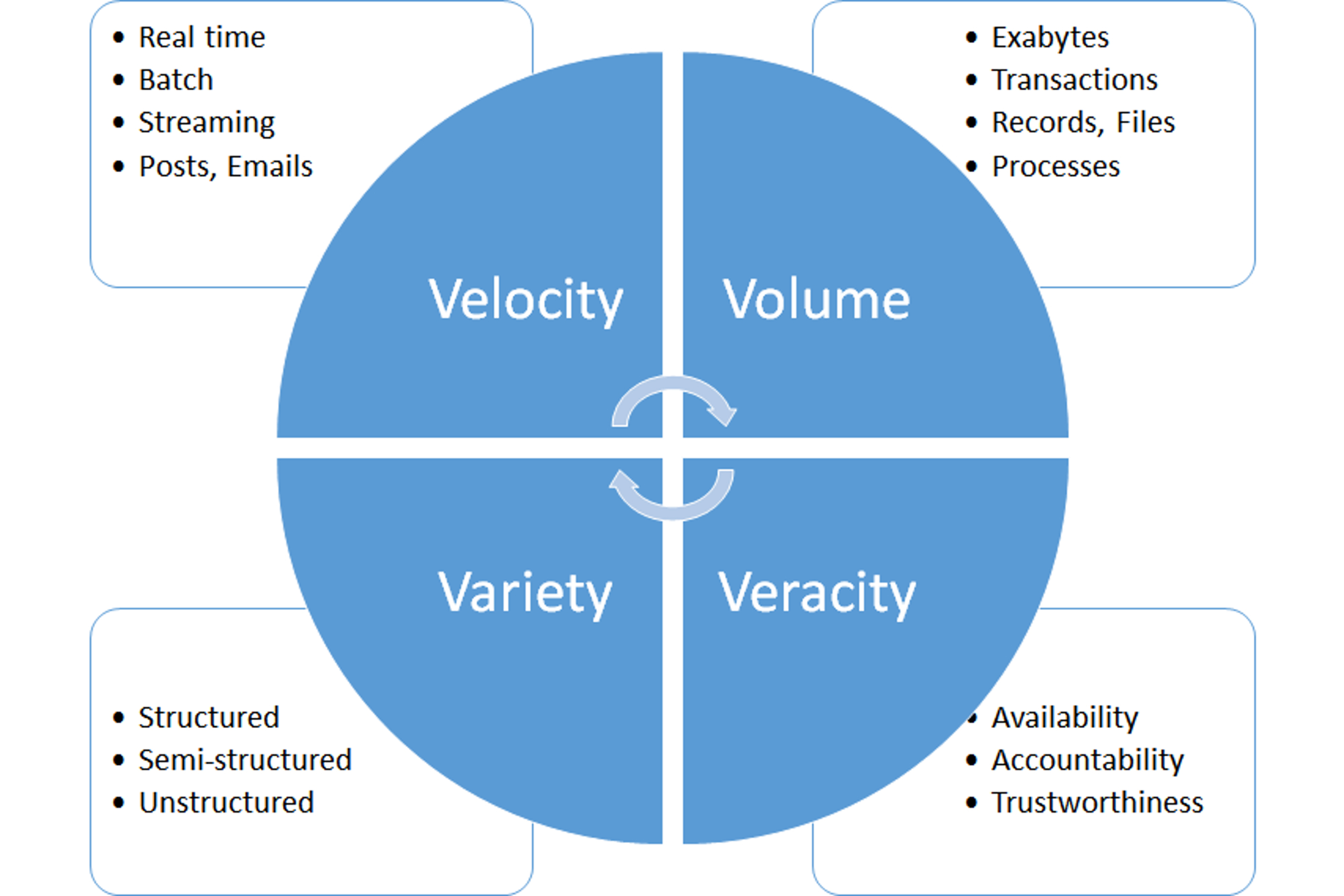}
	\caption{5 Vs of Big Data}
	\label{fig:bigdata}
\end{figure*}

The buzz word of the recent years, Big Data, defined by a number of Vs; Volume, Velocity, Variety, Value and Veracity as shown in Figure \ref{fig:bigdata}.

Volume is the quantity of stored data. Velocity is the speed of data generation or processing. Variety is the type and structure of the data. Value is the importance of information that data provides. Veracity is the variation in quality of data and inconsistency and uncertainty of data.

The survey paper \cite{chen2014big} points the relation between IoT and big data. For example, jet aircraft engines produce one terabyte of data per flight using various sensors. Think about a huge number of flights in a day all around the world and then you can have really big data. HP prepared a business-value white paper related to big data. According to the paper, one trillion sensors, roughly 150 sensors for every person will be existed by 2030. The generated data will be mostly unstructured data and the value of it depends on how the information is extracted from the data. As the number of sensors increases,much more storage and processing will be required. And all of these create some new challenging issues.

Many papers \cite{han2011survey,vicknair2010comparison}, state that the relational database management systems (RDBMS) are inadequate for big data and NoSQL database systems are a solution at least for time being. 

\subsection{NoSQL Graph Databases}
Moniruzzaman et al. \cite{moniruzzaman2013nosql} evaluate NoSQL databases in the aspect of big data analytics. In their survey, they enlist different types of NoSQL databases according to characteristics (features and benefits of NoSQL databases), classification (key-value, document, column-based and graph); and evaluation with a matrix on the basis of few attributes like design, integrity, indexing, distribution, and system. 

There are four types of NoSQL databases which are key-value store (e.g. Amazon’s Simple DB), big table databases (e.g. Apache Cassandra), document-oriented (e.g. MongoDB) and graph databases (e.g. Neo4j). Graph databases \cite{robinson2015graph} consist of nodes and edges (relations between nodes) which store data as properties. Most of graph databases provide the capability to label nodes and edges.
NoSQL Graph databases provide many ways to query data;
\begin{itemize}
	\item User interface via SQL-like query language (Cypher for Neo4j, SQL for OrientDB)	
	\item User interface via Graph visualization to interact with the nodes and edges	
	\item Application program interface (API) to programmatically connect to database
\end{itemize}
Unfortunately, there is not any standardized way of querying, so that you have to write database specific queries every time. There is an open-source framework Apache TinkerPop to provide graph computing capabilities for graph databases. Gremlin is a part of the TinkerPop to traverse the graphs. And by the support of most of the graph databases, any gremlin query can be written once and works on every graph database. 

Graph databases are generally preferred to handle social networks, fraud detection, graph-based operations, real-time recommendations and hierarchical relations.

\section{Related Work}
\label{sec:2}
Over the years, various methods have been used for wireless sensor networks data representation and management (\cite{diallo2012real,bostan2006new}). Our approach is different basically by the aspects of big data, graph database storage, and our unique graph data model.

Yang et al. \cite{li2014wiki} present a service platform which is called Wiki-Health. They have designed platform in three layers; application, query and analysis, and data storage. In data storage layer, they used a NoSQL database as we do. But they use HBase which is a column-oriented key-value store, we use graph database which is better for relational analytics.

\begin{figure*}[!b]
	\centering
	\includegraphics[scale=0.8]{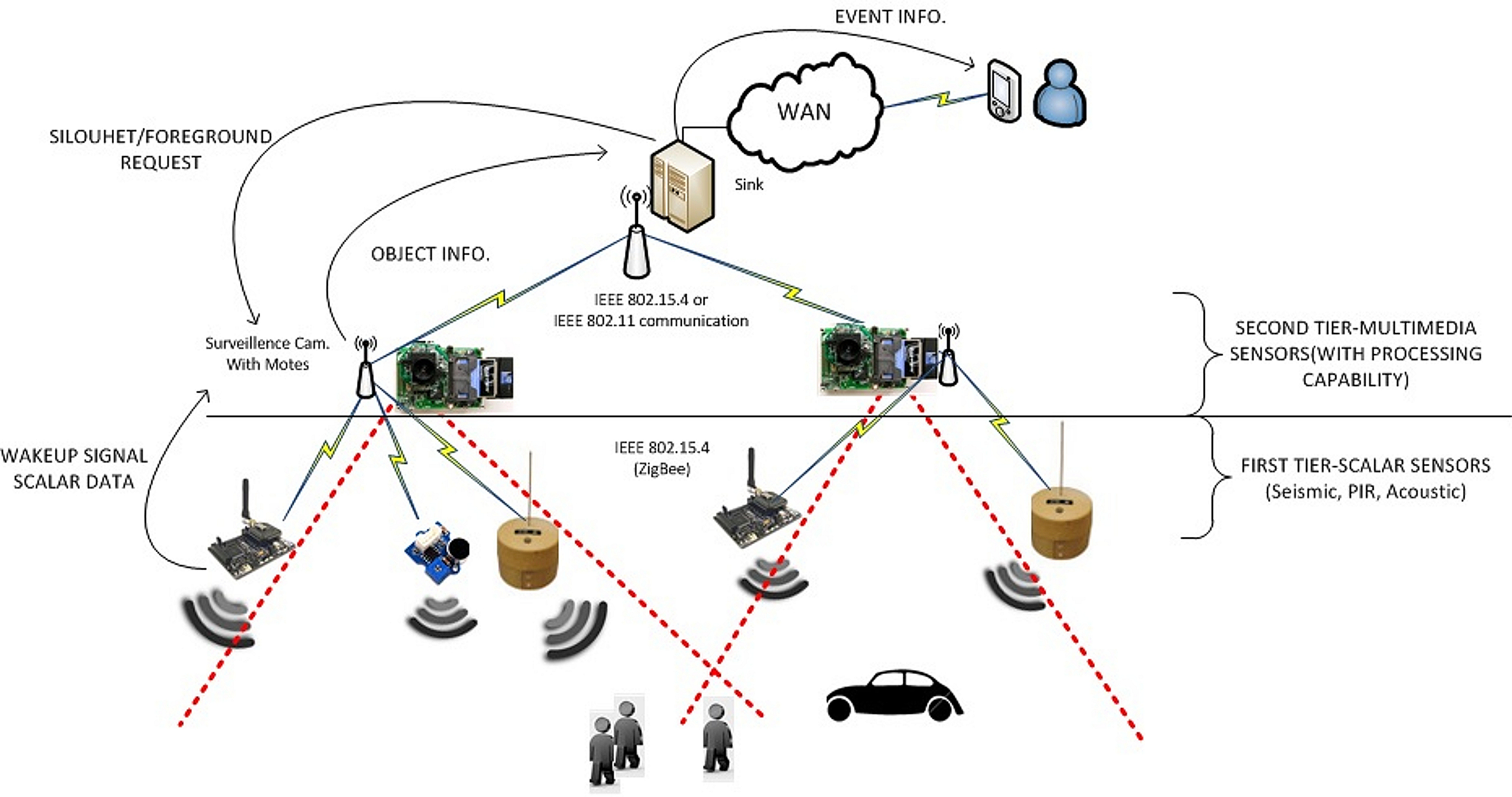}
	\caption{Real WMSN System Architecture}
	\label{fig:realsystem}
\end{figure*}
Christine et al. \cite{jardak2014spatial} discuss big data with spatial data received from wireless sensors using real life scenarios. One of the scenarios is related to smart cities which is similar to surveillance domain. They propose a scalable solution using Hadoop and HBase NoSQL database to prototype a platform for storage and processing wireless data. We also design and implement a simulation prototype from storage layer to analytics layer and more importantly, we propose a grah based data model on top of that architecture. 

Renzo et al. \cite{angles2008survey} present a survey paper on graph database models. They compare graph database models with the other database models, i.e. a relational model. In this paper, we also compare well-known graph database models with  the relational database model. Furthmore we perform queries to benchmark the performance of databases.

Another survey paper is a written by Felemban \cite{felemban2013advanced} which is about border surveillance. His research enlists the literature for experimenting work done in border surveillance and intrusion detection using the technology of WSN. Our research differs from the existing works by employing a graph based approach for surveillance domain and focusing on the simulation of the big data.

PipeNet \cite{stoianov2007pipenet} is a multi-layered wireless sensor networks application focused on pipeline monitoring. System aims to detect the leaks and other anomalies in water pipelines. They have used various types of sensors like pressure, pH and ultrasonic sensors on top of Intel Moto platform. Our sensor nodes are built on Rasberry Pi platform and have seismic, acoustic, and PIR sensors but also a multimedia camera. A camera needs further analysis like image processing and feature extraction. Their multi-layer architecture is similar to our prototype but in another domain with different sensors and different analytical approaches. They try to analyze the collected multi-modal data for detection of the leaks. But we try to identify objects and track their movement. On the other hand, we approach our sensor data as big data.

Suvendu Kumar et al. \cite{mohapatra2016big} propose an analytic architecture for big data to detect intruders using camera sensors. Our work differs by using additional scalar sensors like acoustic and seismic sensors but also proposed graph based data model.

\section{Real WSN System}
Our reference WMSN system is composed of wireless multimedia sensors and some scalar sensors. The system is designed as a multi-tier automated surveillance system. The first layer is the sensing layer with scalar sensors including acoustic, seismic, and PIR. The second layer is triggered by first layer. Multimedia sensors like camera and microphone are used capture video and audio. After applying the fusion, object type and location of the sensor is extracted to be provided to the next layer, which is called the sink layer. The sink layer provides the capability to do analytics on all collected and generated information in the network.

The overall architecture is given in Figure \ref{fig:realsystem}. Sensor nodes are connected to the gateway nodes via ZigBee (IEEE 802.15.4) interfaces. A special thread for serial messaging is developed to send the events from sensor nodes to the gateway. The gateway prepares XMPP messages using the gathered events coming from leading nodes via broadcast messages. Those XMPP messages are transferred to the sink node. The XMPP messages and multimedia data is transfered over IP based (IEEE 802.11) connection.

The hardware of our sensor node is based on a Raspberry Pi (RPi) 512-MB Model B board and includes the following hardware components:
\begin{itemize}
	\setlength\itemsep{0pt}
	\item ARM1176 700MHz processor,
	\item Graphical processing unit (GPU),
	\item 512 MB SDRAM shared with GPU,
	\item SD card slot for on board storage,
	\item On board 10/100 Mb Ethernet port,
	\item 2x USB 2.0 ports,
	\item 1 CSI input connector for the camera module,
	\item Video and audio outputs,
	\item GPIO ports,
	\item 5V 700-mA microUSB power requirement.
\end{itemize}
The following list is the components installed on a node in our WSN system to fulfill its functions:
\begin{itemize}
	\setlength\itemsep{0pt}
	\item Motion sensor (PIR),
\item Acoustic sensor (AS),
\item Vibration sensor (VS),
\item Raspicam camera module,
\item Xbee ZigBee (IEEE 802.15.4) adapter,
\item 4400-mAh 5V 1A power bank,
\item Microphone,
\item Wi-Fi (IEEE 802.11) dongle,
\item XMPP client software
\end{itemize}

Sensor to sensor communication, as well as gateway to sink communication, is completed via ZigBee interfaces. Nodes are equipped with low-bandwidth radio devices using IEEE 802.15.4 (ZigBee) standard and ZigBee provides a line of sight up to of 1500 m. at outdoor conditions and 250 Kbps at most.

In our real WSN system, sensor node and gateway roles are all predefined, there is not any dynamic gateway selection. Because different roles may need different kinds of hardware components.
\section{The Graph-Based Big Data Model}
\label{sec:3}
Our graph-based big data model is built on the multimedia sensor networks topology. There is a sink node in the base station and there are a number of clusters connected to the sink. Each cluster consists of a gateway, which can be called the cluster head, and a set of sensor nodes.

The data flow occurs from the sensor nodes to gateways and from gateways to the other gateways (multi hop) and finally to the sink node. Each sensor node holds a set of data sensed by the sensors and camera of the node. Sensor nodes include embedded programs for handling correlation, transformation, and aggregation on the raw data, which is called first level fusion. The sensor fused data are reported to the leading gateway by all of its connected sensor nodes.

The gateway waits for all sensor nodes to report. When all reports are ready, the gateway applies an aggregation or filtering on the received data. The second-level fusion is done at this point and the output of the fusion is a summary of that cluster. The gateway fused data are forwarded to the sink node for a final decision.

\begin{figure}[!b]
	\includegraphics[scale=0.9]{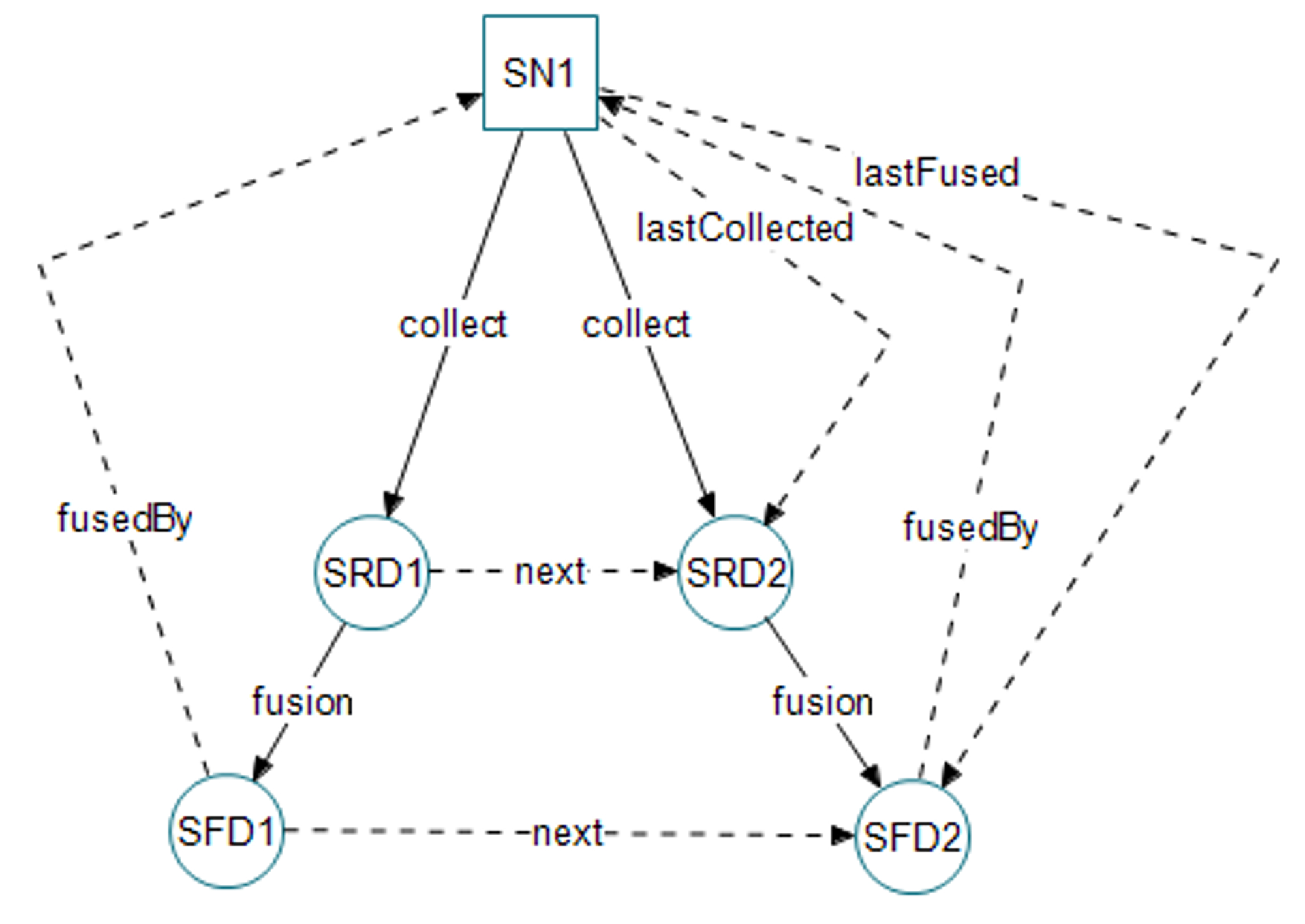}
	\caption{First Level Fusion model (SN:Sensor Node, SRD:Sensor Raw Data, SFD:Sensor Fused Data)}
	\label{fig:3}
\end{figure}

Similar to the second-level fusion, the sink waits for all gateways’ fused data. By applying some patterns to detect anomalies or other kinds of analysis are done at the third level fusion. The output of the last level fusion is an action like triggering an alarm or a notification message to another system.

Figure \ref{fig:3} shows the first level fusion graph model from raw data collection to fusion. The sensor node is responsible for fusion at this level. The input is the raw data and the output is sensor fused data. Last collected raw data and last fused fusion data are explicitly pointed for the purpose of the direct link.

\begin{figure}
	\includegraphics[scale=0.85]{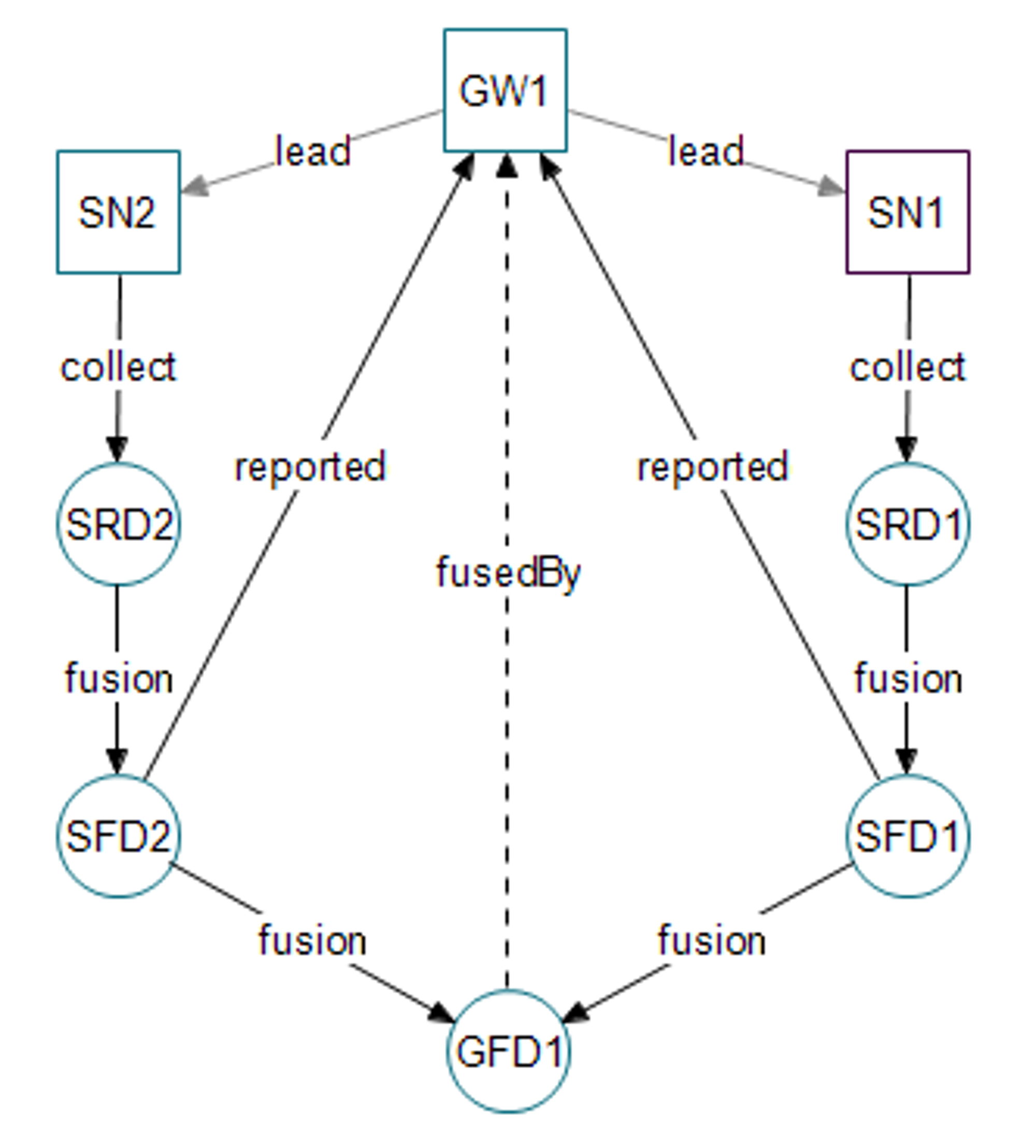}
	\caption{Second Level Fusion model (GW:Gateway, SN:Sensor Node, SRD:Sensor Raw Data, SFD: Sensor Fused Data, GFD:Gateway Fused Data)}
	\label{fig:4}
\end{figure}

Figure \ref{fig:4} shows the second level fusion graph model to identify the relations between sensor nodes and the gateway. The gateway is responsible for the fusion at this level. The input is all sensor fused data coming from sensor nodes and the output is gateway fused data which can be filtered, aggregated or transformed data.

\begin{figure}
	\centering
	\includegraphics[scale=0.85]{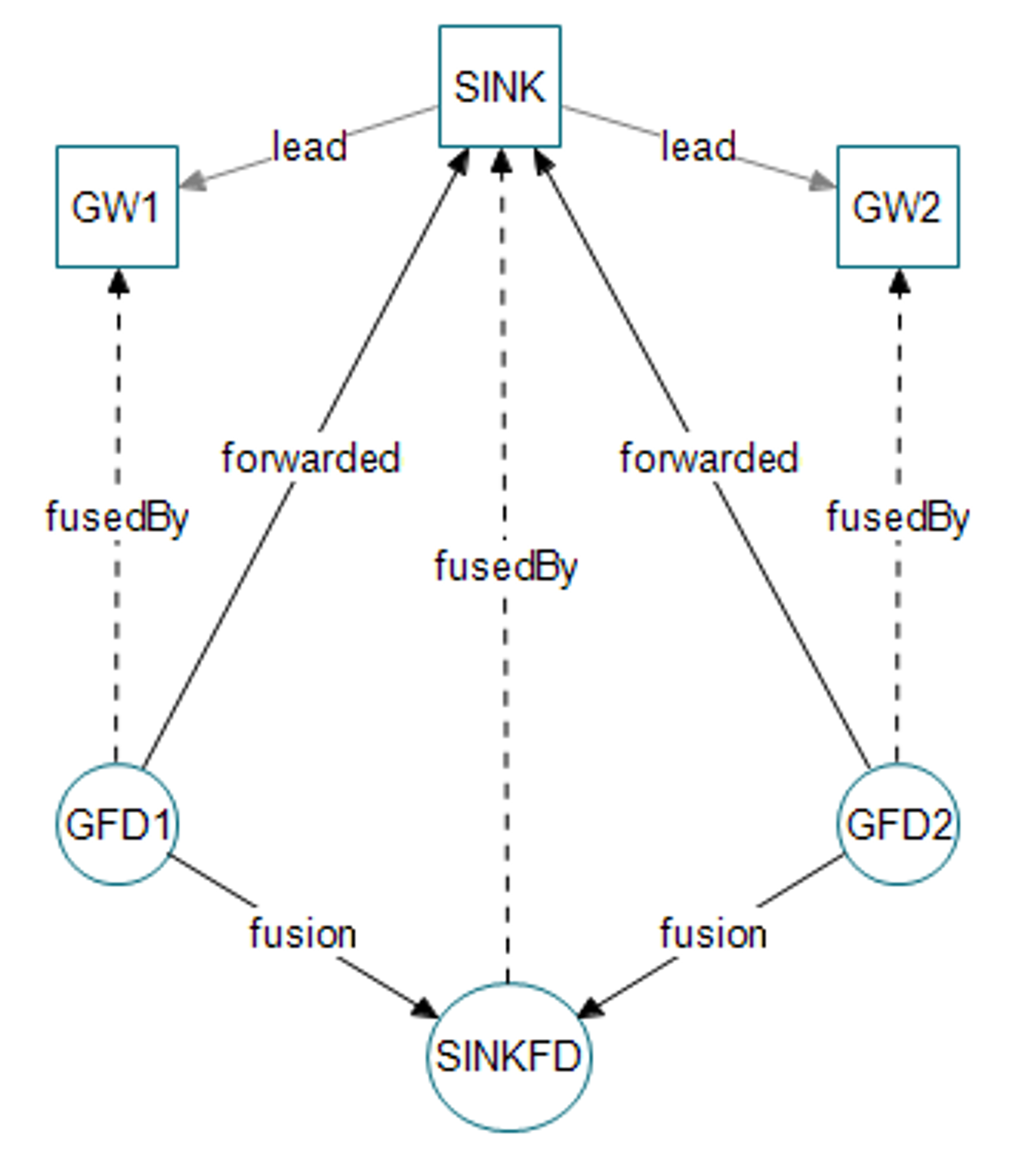}
	\caption{Third Level Fusion model (GW:Gateway, GFD:Gateway Fused Data, SINKFD:Sink Fused Data)}
	\label{fig:5}
\end{figure}

Figure \ref{fig:5} shows the third level fusion graph model which is executed at the sink to aggregate all data fused from gateways. The input is the all fused data coming from gateways and the output is sink fused data which cause actions like triggering an alarm or notifying the operators.

\section{Prototype Implementation for Proposed Model}
\label{sec:4}
We have already implemented the proposed graph model using OrientDB Graph Database and developed a simulator to generate synthetic data. In the following subsection, we describe why OrientDB graph database is chosen as default storage system. Other subsections show the implementation of the data model with the detailed design of generic infrastructure and simulation infrastructure.

\subsection{Graph Database Selection}
Our research includes applying the currently available databases to our graph-based big data model. Salem et al. \cite{hadim2006middleware} compare a set of databases like Cougar and TinyDB. Li-Yung Ho et al. \cite{ho2012distributed} propose a distributed graph database based on an open-source graph database which is called Neo4j. The options are limited if you are looking for a graph database. Neo4j, Titan, and OrientDB are featured open-source graph databases.

Neo4j is a well-known graph database and used by many researchers. In addition, Neo4j is relatively easier to be used rapidly by developing some small pieces of code. Spring Framework support is really helpful to put things together very fast.

Titan is another open-source option for a graph database but its development is stopped and discontinued in early 2015. Therefore we did not prefer to utilize Titan.

OrientDB is another open-source graph database which is not as popular as Neo4j for now but has many advantages over it. Table \ref{tab:compare} shows the comparison of OrientDB and Neo4j Community Editions which is provided by the official  website of OrientDB. From all those compared features, "Multi-Master Replication", "SQL" and "Elastic Scalability with Zero Configuration" are the most important features for us to choose OrientDB.

\begin{table}
	\caption{Compare OrientDB and Neo4j Community Editions}
	\label{tab:compare}
	\renewcommand{\arraystretch}{1.1} 
	\begin{tabular}{@{}lll}	
		\hline
		{Feature} & {OrientDB} & {Neo4j} \\
		\hline
		Graph Database	&	Yes	&	Yes	\\   
		TinkerPop Standard Compliance	&	Yes	&	Yes	\\    
		ACID Transaction	&	Yes	&	Yes	\\    
		Unique Constraints	&	Yes	& Yes		\\ 
		Fulltext Support	&	Yes	& Yes		\\  
		Spatial Support	&	Yes	& Yes		\\   
		Java Hooks	&	Yes	& Yes		\\       
		Record Level Security	&	Yes	&	No	\\  
		User and Role Security	&	Yes	&	No	\\  
		SQL	&	Yes	&	No	\\  
		Dynamic Triggers	&	Yes	&	No	\\  
		Custom Data Types	&	Yes	&	No	\\  
		Additional Constraint Types	&	Yes	&	No	\\  
		Indexes on Multiple Properties	&	Yes	&	No	\\  
		Different Schema Modes	&	Yes	&	No	\\  
		Multi-Master Replication	&	Yes	&	No	\\  
		Sharding	&	Yes	&	No	\\  
		Elastic Scalability	&	Yes	&	No	\\  
		Server-Side Functions	&	Yes	&	No	\\  
		Embeddable with No Restrictions	&	Yes	&	No	\\  
		Sequences	&	Yes	&	No	\\  
		\hline
	\end{tabular}
\end{table}

\subsection{Data Model Implementation}

Nodes (vertices) and relations (edges) are defined in graph databases to store data. Compared to the traditional RDBMS approach, every row in a table is replaced with a node and its properties. There are edges to represent cross-table references.

At the first step, we define the node and edge types. Node types are; Sink, Gateway, SensorNode, SensorRawData, SnFusedData (\textit{SensorFusedData}), GwFusedData (\textit{GatewayFusedData}) and SinkFusedData. Edge types are; Lead, Collect, LastCollected, Next, Fusion, FusedBy, LastFusion, Reported and Forwarded. The edge types are defined for the usage between specific nodes. Table \ref{tab:1} lists the edge types in our graph database.

Each sensor node has the capability to hold temporarily a set of data which are sensed by sensors like \textit{PIR}, \textit{seismic}, \textit{acoustic} and camera. That capability is provided by an in-memory database. For the fusion at the sensor node level, this in-memory database is used as a cache to analyze the changes in scalar sensors and provide some additional data to the first level fusion. The algorithm of the first level fusion is shown in Algorithm \ref{alg:1}. The fusion result is stored in the \textit{fusedData} including the \textit{video}, \textit{silhouette}, \textit{foreground} and low level \textit{features}.

\begin{algorithm}
	\label{alg:1}
	\caption{Sample first level fusion algorithm executed on sensor nodes}
    \SetKwInOut{Input}{Input}
    \SetKwInOut{Output}{Output}
    \SetKwProg{Fn}{Function}{}{end}
	\Fn{firstLF($PIR, seismic, acoustic, threshold$)}{
	    \Input{Boolean $PIR$ identifies if there is a movement or not, two integers $seismic$ and $acoustic$ values are scalar data sensed by the node,  $threshold$ is used to identify that there is an object with enough sound and vibration on sensor node}
	    \Output{$fusedData$ is the fusion data involving the multimedia data}
	    
	    initialize $fusedData$
	       
	    \If{$PIR=true$ and $seismic \ge THRESHOLD$ and 	$acoustic \ge threshold$}
		{
		    $fusedData.video$ = startVideoRecording() \;          
		    $fusedData.frame$ = selectFrame($fusedData.video$)\;        
		    $fusedData.frmFeatures$ = findLowLevelFeatures($fusedData.frame$)\;
		    $fusedData.foreground$ = selectForeground($fusedData.frame$)\;
		    $fusedData.fgndFeatures$ = findLowLevelFeatures($fusedData.foreground$)
		    
		    $fusedData.silhouette$ = extractSilhouette($fusedData.foreground$)\;
		}
	  
	  	return $fusedData$;
	}
\end{algorithm}

\begin{table} 
	\caption{Node and Edge Types}
	\label{tab:1}
	\renewcommand{\arraystretch}{1.1} 
	\begin{tabular}{@{}lll}	
		\hline
		Edge Type & From Node & To Node \\
		\hline
		Lead	&	Gateway	&	SensorNode	\\   
		Lead	&	Gateway	&	Gateway	\\    
		Lead	&	Sink	&	Gateway	\\    
		Collect	&	SensorNode	& SensorRawData		\\ 
		LastCollected	&	SensorNode	& SensorRawData		\\  
		LastFusion	&	SensorNode	& SnFusedData		\\   
		LastFusion	&	Gateway	& GwFusedData		\\       
		Next	&	SensorRawData	&	SensorRawData	\\  
		Next	&	SnFusedData	&	SnFusedData	\\  
		Fusion	&	SensorRawData	&	SnFusedData	\\  
		Fusion	&	SnFusedData	&	GwFusedData	\\  
		Fusion	&	GwFusedData	&	SinkFusedData	\\  
		FusedBy	&	SnFusedData	&	SensorNode	\\  
		FusedBy	&	GwFusedData	&	Gateway	\\  
		FusedBy	&	SinkFusedData	&	Sink	\\  
		Reported	&	SnFusedData	&	Gateway	\\  
		Forwarded	&	GwFusedData	&	Gateway	\\  
		Forwarded	&	GwFusedData	&	Sink	\\  
		\hline
	\end{tabular}
\end{table}

Sensor nodes apply the first level fusion and reports the output \textit{fusedData} to leading gateway. The gateway applies the second level fusion as in Algorithm \ref{alg:2}. The purpose of fusion at the gateway is a refinement of the sensor fused data before sending to the sink node. As a sample refinement algorithm, removing the duplications and normalizing the data according to scalar data can be written.

\begin{algorithm}
	\label{alg:2}
	\caption{Sample second level fusion algorithm executed on gateways}
	\SetKwInOut{Input}{Input}
	\SetKwInOut{Output}{Output}
	\SetKwProg{Fn}{Function}{}{end}
	\Fn{secondLF($fusedDataList, threshold$)}{
		\Input{The list of $fusedData$ reported by leading sensor nodes, $threshold$ is used to identify the difference between two scalar value}
		\Output{Filtered list of $fusedData$ by duplication removal}
		
		initialize an empty array $filteredDataList$
		
		\For{$i = 0$ \KwTo $fusedDataList.size$}{
			$current$ = $fusedDataList[i]$\;	
			$previous$ = previous element of $current$\;
			
			$diff$ = $current.acoustic$ - $previous.acoustic$\;
			
			$diffRate$ = $current.acoustic$ * $threshold / 100$\;
			
			\eIf{$diff < diffRate$}
			{
				mark $current$ as duplicate and drop
			}
			{
				add $current$ to $filteredDataList$
			}
		}	
		return $filteredDataList$
	}
\end{algorithm}

The output of the second level fusion is reported to the sink node. As the final decision maker, the sink correlates the concepts forwarded by gateways and decides that if an \textit{action} is necessary or not. If an action is required, notification of the operator is triggered as given in Algorithm \ref{alg:3}. 
All those three fusion algorithms are sample algorithms to provide the proof of concept execution of all phases of the simulated environment.

\begin{algorithm}	
	\label{alg:3}
	\caption{Sample third level fusion algorithm executed on the sink}
	\SetKwInOut{Input}{Input}
	\SetKwInOut{Output}{Output}
	\SetKwProg{Fn}{Function}{}{end}
	\Fn{thirdLF($filteredDataList, threshold$)}{
		\Input{The list of $fusedData$ reported by reporting gateways, $threshold$ is used to identify the difference between two scalar value}
		\Output{Actions generated by reported $fusedData$ list}
		
		initialize an empty array $actionList$
		
		\For{$i = 0$ \KwTo $filteredDataList.size$}{
			$current$ = $fusedDataList[i]$\;	
			$previous$ = previous element of $current$\;
			
			\If{$current.acoustic > threshold$ and $previous.acoustic > threshold$}
			{
				$action$ = createAction($current$)\;				
				add $action$ to $actionList$\;
				notify operator using $action$\;
			}
		}	
		return $actionList$
	}
\end{algorithm}

All those three fusion algorithms are sample algorithms to provide the proof of concept execution of all phases of the simulated environment.

\subsection{Generic Infrastructure}
We have developed the whole system by using Java 1.8 as maven projects to use Apache Maven as the dependency management framework. To develop the application of our research project beyond OrientDB, we have designed a generic infrastructure which can be easily adopted to other database systems. To achieve that, we have developed business logic without any dependency to OrientDB. 

There are two managers called DataManager and NetworkManager. DataManager defines the necessary interfaces to populate data for the underlying database. NetworkManager has the business logic to construct network topology and defines the necessary interfaces to create network entities for the underlying database.

As there is a data flow between sensor nodes and gateways and sink, in order to cope with the bottleneck of high throughput of streaming data, we position Apache ActiveMQ message queues between each process. We define message queues as seen on Table \ref{tab:mq} below. Fusion logic is developed on top of messaging queues and there are 3 business logic handlers; Level1and2Fusion, GatewayForward and Level3Fusion.

\begin{table}
	\caption{Defined Message Queues}
	\label{tab:mq}
	\renewcommand{\arraystretch}{1.1} 
	\begin{tabular}{@{}lll}		
		\hline
		{Queue Name} & {Process} & {Queue Element} \\
		\hline
		Scalar Data	&		Collected Raw Data	&		SensorRawData	\\   
		Fused	&		Level 1 and 2 Fusion		&	SnFusedData	\\   
		Forward	&		Forwarding to Sink		&	GwFusedData	\\   
		Action	&		Level 3 Fusion	&		SinkFusedData	\\   
		\hline
	\end{tabular}
\end{table}

\section{Simulation}
\label{sec:6sim}
To develop and experiment the graph data model, we have developed a simulator which is mainly focused on data simulation but also supports network topology simulation. 
\subsection{Network Topology Simulation}
\begin{figure}[!b]
	\centering
	\includegraphics[scale=0.5]{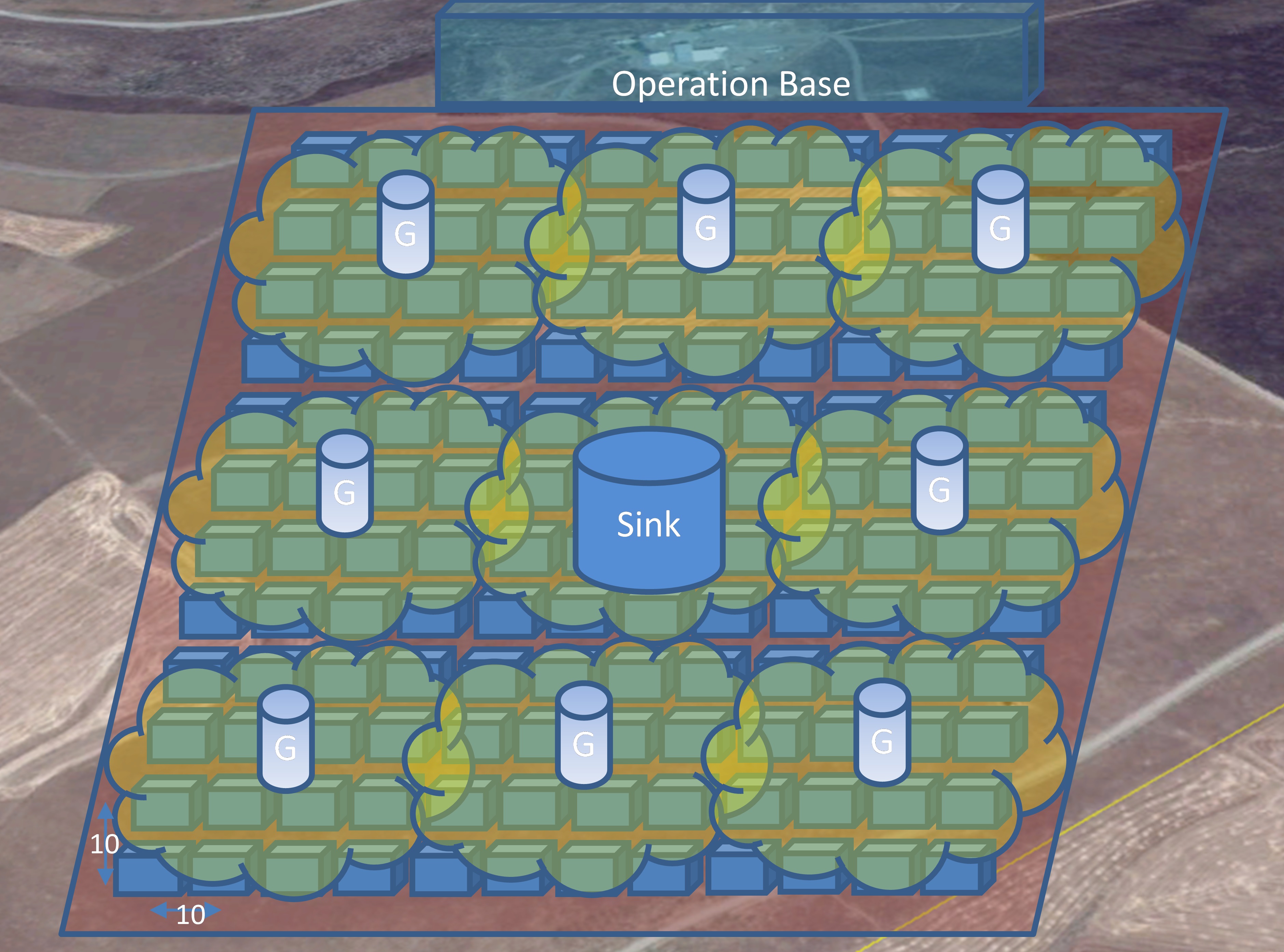}
	\caption{Simulated Network Topology}
	\label{fig:map}
\end{figure}

For our network topology, we assume that nodes are distributed with a grid layout to the simulation area which is assumed to be square. Figure \ref{fig:map} shows a visualization of 16 sensor nodes in each cluster with a gateway in the middle and 9 gateways in total with Sink in the center. Between two sensor nodes, the distance is 10 units. Each cluster has 4 sensor nodes at one side and there are 3 clusters in an edge of the total area. So, we have 120x120 sized grid layout for our simulation.

\subsection{Data Flow Simulation}
Data flow simulation is started with Raw Data Generator which produces synthetic data with position, PIR, seismic and acoustic information. These data are sensed by sensor nodes which are close to the position. A sensor raw data object is created for each sensor node with the calculated sensor data according to the distance to the position of the raw data created by Raw Data Generator. Then the generic infrastructure explained in the previous section is used to simulate all processes of the data flow.

\begin{figure}
	\centering
	\includegraphics[scale=0.54]{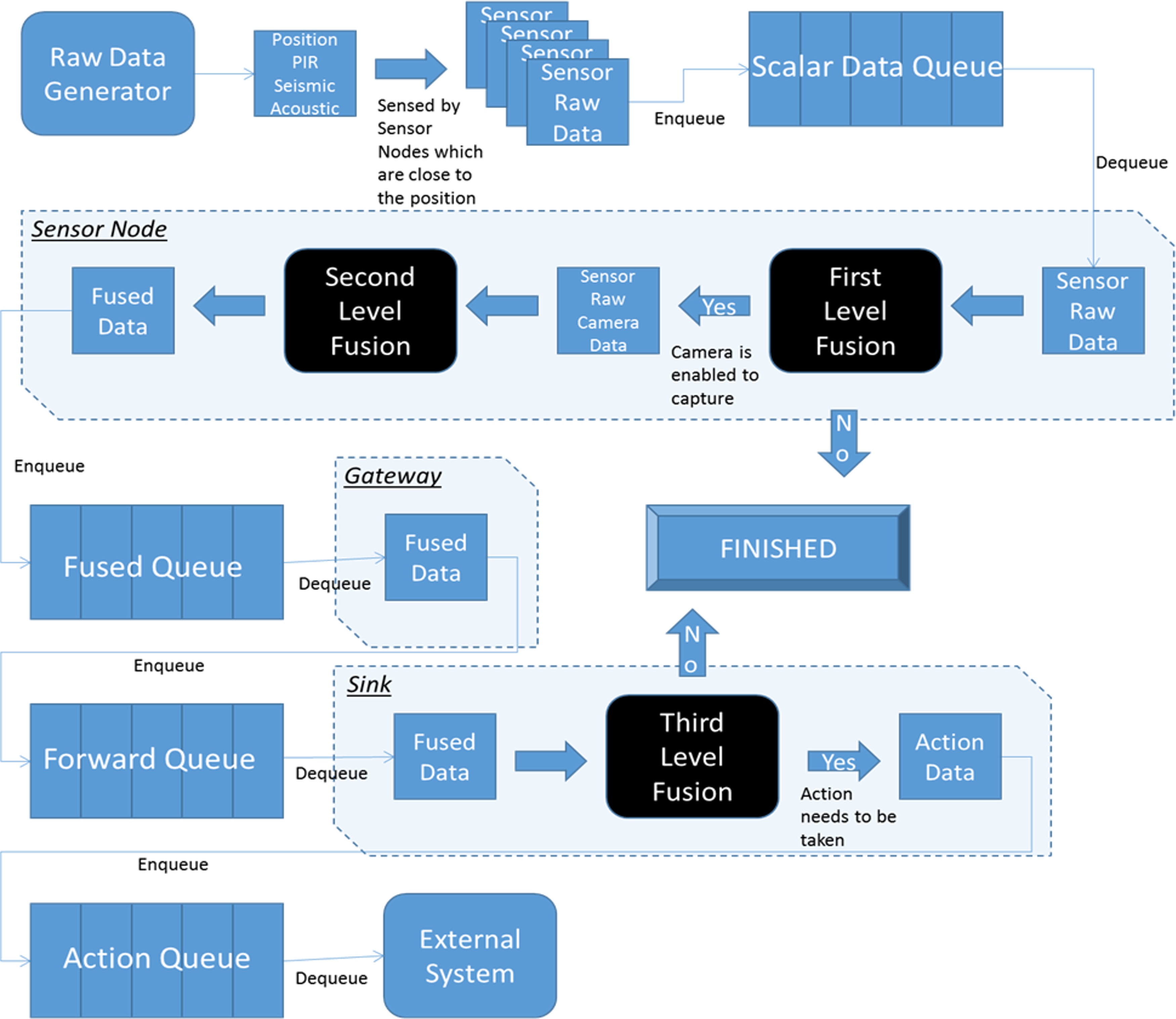}
	\caption{Data Flow Simulation}
	\label{fig:df}
\end{figure}

\subsection{Simulator}
Currently, we adopt our simulation infrastructure on OrientDB, Neo4j which is the main rival of OrientDB and MySQL for the relational to graph database comparison. Every simulation is replicated to all three databases simultaneously so that the same test environment is provided.

\begin{figure}
	\centering
	\includegraphics[scale=0.95]{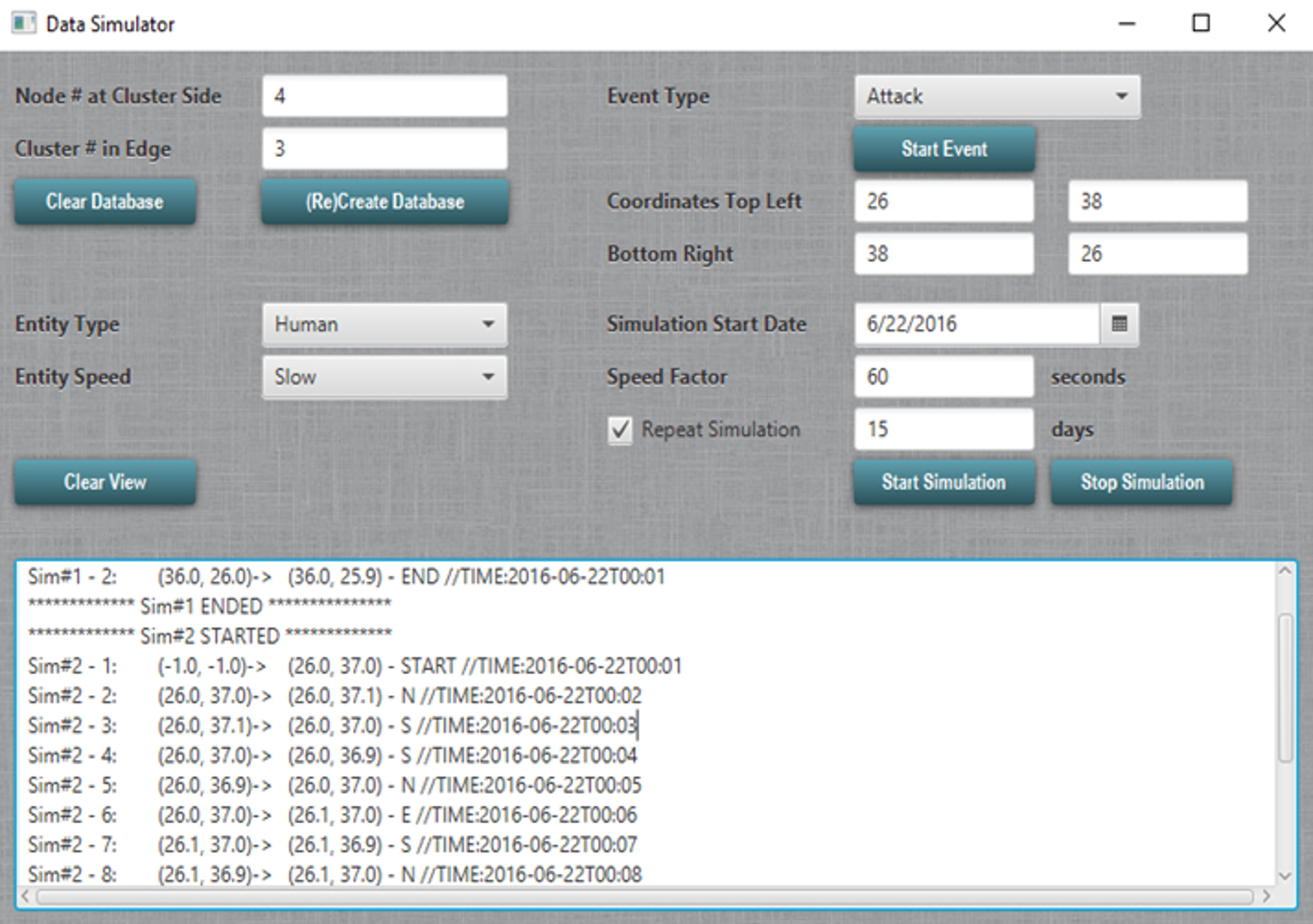}
	\caption{Data Simulator}
	\label{fig:sim}
\end{figure}

Figure \ref{fig:sim} is the screenshot of our data simulator to generate the network topology in Figure \ref{fig:map}. “(Re)Create Database” button cleans the database and generates sink, gateways and sensor nodes according to the given parameter related to node count and cluster count.

“Start simulation” button starts simulation by sending an entity from one of the edges of the area. Then, the entity moves randomly according to its speed and simulation ends when the entity moves out of the area. Possible moves are going to north, south, east, or west and don’t move. 

It is possible to run parallel simulations by clicking the button at any time. And if “Repeat Simulation” checkbox is checked, a new simulation is automatically started when current simulation ends.

There are several Entity types to simulate. Each entity has its own speed, acoustic and seismic values. Table \ref{tab:entities} show each type and its simulation values. “Entity Speed” selection combobox decreases or increases the speed value of the simulating entity. Acoustic and seismic values are defined by the selection of "Entity Type" combobox and according to entity’s distance from the sensor node, those values are recalculated to degrade its effect on the sensor node.

\begin{table}[!b]
	\centering
	\caption{Simulated Entity Types}
	\label{tab:entities}
	\renewcommand{\arraystretch}{1.1} 
	\begin{tabular}{llll}	
		\hline
		{Entity Type} & {Speed} & {Acoustic} & {Seismic} \\
		\hline 
		Human &	1 &	20 &	10 \\
		Animal &	2 &	40 &	20 \\
		Vehicle &	4 &	70 &	80 \\
		GroupOfHuman &	1 &	60 &	30 \\
		GroupOfAnimal &	2 &	80 &	60 \\
		\hline 
	\end{tabular}
\end{table}

Assume that simulation put an “Animal” entity at (37,120) position. The entity is between the sensor nodes at (30,120) and (40,120). Table \ref{tab:node30} and Table \ref{tab:node40} show the sensed values between those two nodes for Animal Type. 

\begin{table*}

	\caption{Sensed Values at Node(30,10)}
	\label{tab:node30}
	\renewcommand{\tabcolsep}{0.6pc} 
	\renewcommand{\arraystretch}{1.1} 
	\begin{tabular}{lllllllllllllll}	
		\hline
		{ANIMAL} & {30,10} & {31,10} & {32,10} & {33,10} & {34,10} & {35,10} & {36,10} & {37,10} & {38,10} & {39,10} & {40,10} \\
		\hline 
		Acoustic	&	20	&	18	&	16	&	14	&	12	&	10	&	8	&	6	&	4	&	2	&	0	\\ 
		Seismic	&	40	&	36	&	32	&	28	&	24	&	20	&	16	&	12	&	8	&	4	&	0	\\ 
		\hline 
	\end{tabular}
\\
\\

	\caption{Sensed Values at Node(40,10)}
	\label{tab:node40}
	\renewcommand{\arraystretch}{1.1} 
	\begin{tabular}{lllllllllllllll}	
		\hline
			{ANIMAL} & {30,10} & {31,10} & {32,10} & {33,10} & {34,10} & {35,10} & {36,10} & {37,10} & {38,10} & {39,10} & {40,10} \\
			\hline 
		Acoustic &	0 &	2 &	4 &	6 &	8 &	10 &	12 &	14 &	16 &	18 &	20	\\ 
		\hline 
		Seismic &	0 &	4 &	8 &	12 &	16 &	20 &	24 &	28 &	32 &	36 &	40	\\ 
		\hline 
	\end{tabular}
\end{table*}

Another important capability of the simulation tool is Event generation. We can identify event types and generate it at any time while the simulation is running. At first stage, we have 2 types of events. 
\begin{itemize}
	\item Attack: As seen in Figure \ref{fig:map}, operational base is located at north. If something comes from south and directly moves toward the base, this movement is an Attack event for us.
	\item Smuggling: If a group of human is moving together with a group of animal and they are coming from west and going in the direction of south-east, this movement is smuggling event for us.
\end{itemize}
When “Start Event” button is pressed, selected event is started to be simulated. Additional event types can be added for further analysis.

\section{A Case Study: Surveillance Application}
\label{sec:5}
Surveillance systems need robust and scalable infrastructure. To achieve that, all data flow and data itself are needed to be analyzed and modelled.

A set of sensors and video cameras are needed to monitor the whole city. Assume that, we cluster the sensors and cameras according to the districts and each cluster forwards sensed data to the HQ (Head Quarter) which is the operation center. At the HQ, an alarm is triggered, or a notification is sent to the officers to early detection of violence.

Sensor types can be seismic, acoustic and PIR (Passive Infrared) which are types of scalar sensors. In addition to them, video cameras or thermal cameras can be added to critical locations. As the default, cameras are switched off. According to the sensed information from scalar sensors, the predefined conditions can be extracted using rule-based approaches. The motion or environmental change may be detected and interpreted to activate the camera by providing a rough prediction of the moving object.

We categorize the moving object as; Animal, Human, and Vehicle. The data collected from scalar sensors are analyzed to guess the category of the objects according to predefined thresholds (Table \ref{tab:2}).

\begin{table}[!b]
	\caption{Sample Thresholds To Identify Objects}
	\label{tab:2}
	\renewcommand{\arraystretch}{1.1} 
	\begin{tabular}{llll}	
		\hline
		{Object Type} & {PIR} & {Seismic (Hz)} & {Acoustic (dB)} \\
		\hline
		Animal & True	&	5 - 20	&	5 - 30	\\   
		Human	& True & 21 - 55	&	31 - 50	\\    
		Vehicle	&	True & \textgreater 35	& \textgreater 50 \\   
		\hline
	\end{tabular}
\end{table}

After activation of the camera by analyzing scalar sensor data, video and audio streams from the camera are started to be processed. That processing in the sensor node is called the first level fusion. After fusion, we have a concept of the moving object and maybe even a silhouette. Fusion output is reported to the gateway which is the leading node of that district.

A gateway is connected to a set of sensors and cameras. The described operations are done for all leaded sensors so that the gateway receives many concepts and silhouettes. By applying some algorithms like filtering the duplicate concepts or aggregating them to normalize the received data, the gateway accomplishes the second level fusion. After fusion, we have more accurate concepts and silhouettes provided by many sensors. Fusion output is forwarded to the HQ (Headquarter) for a final decision.

As there are many districts in a city, there are many gateways which are far away from the HQ and not directly connected to it. In that case, data is forwarded over other gateways. At the HQ, the received data from all districts are analyzed, aggregated or filtered for the purpose of detecting anomalies or finding some patterns, which is called the third level fusion. At the end, the fusion comes to the conclusion and an alarm is triggered to the officers or the external system of the armed forces is notified.

\section{Experimental Work}
\label{sec:6}
We setup a test environment to make some experiments on our graph model. Test environment specifications are;
\begin{itemize}
	\item Intel i7-4710HQ Quad Core CPU
	\item 16 GB DDR3 RAM
	\item 240 GB SSD Storage
	\item 4 GB NVIDIA 860GTX GPU
\end{itemize}
Test environment has three database systems which are;
\begin{itemize}
	\item OrientDB v2.2.5 (Graph database)
	\item Neo4j v2.3.2 (Graph database)
	\item MySQL v5.7.1 (Relational database)
\end{itemize}

We have simulated a multimedia wireless sensor network with synthetic raw data. Sensor nodes are placed in a square shaped area. Gateways are located in the center of each group of sensor nodes.  The sink node is placed in the center of the whole area as shown in Figure \ref{fig:map}.

There are 25 million sensor nodes which are leaded by 2,500 gateways. There is sink node to collect all data in our simulation environment. Each gateway leads 10,000 sensor nodes.

\subsection{Comparison to Relational Data Model}
\begin{figure}
	\centering
	\includegraphics[scale=0.9]{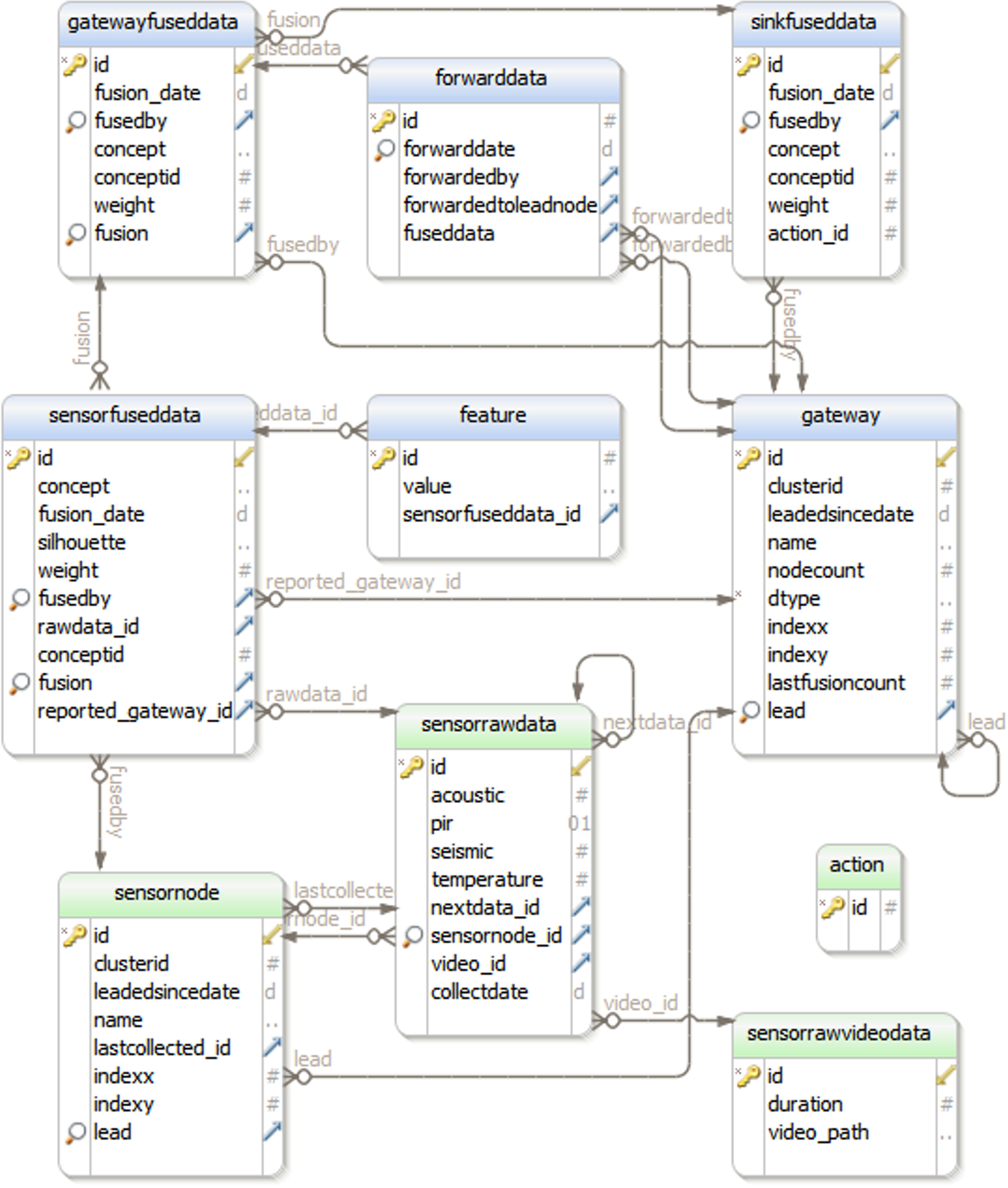}
	\caption{Relational Database Schema}
	\label{fig:7}
\end{figure}

This experiment is done on all three databases installed in our test environment. We have run many queries on both our graph data model and relational data model. Figure \ref{fig:7} shows the relational database representative of our graph data model on MySQL.

Below queries are randomly selected sample queries and Table IV shows the test results of the experiment.
\subsubsection{Concepts Based Query}
This query finds the specific type of objects with the highest probability of its detection time and location.

\vspace{5pt}
\textit{OrientDB Query:}
\begin{lstlisting}
SELECT concept, weight, fusionDate, out("fusedby").indexX, out("fusedby").indexY FROM fuseddata WHERE weight>0.90 AND concept = "Vehicle" ORDER BY fusionDate
\end{lstlisting}	
\hspace{5pt}
\textit{Neo4j Query:}
\begin{lstlisting}
MATCH (b:fuseddata)-[:fusedby]->(sd:sensornode) WHERE b.concept = "Human" AND b.weight>0.90 RETURN b.concept, b.weight, b.fusionDate, sd.indexX, sd.indexY
\end{lstlisting}
\hspace{5pt}	
\textit{Relational SQL Query:}
\begin{lstlisting}
SELECT b.concept, b.weight, a.fusion_Date, sd.indexx, sd.indexy FROM sinkfuseddata a, gatewayfuseddata g, sensorfuseddata b, sensornode sd WHERE a.concept = 'Vehicle' AND a.weight>0.90 AND a.id = g.fusion AND  g.id = b.fusion AND b.fusedby = sd.id ORDER BY a.fusion_Date ASC
\end{lstlisting}

\subsubsection{Video Based Query}
This query finds the possible explosions by identifying continued high volume around the surveillance area with their recorded video paths and video duration. The value bigger than 15 is assumed to be a high volume sound.
	
\vspace{5pt}
\textit{OrientDB Query:}
\begin{lstlisting}
SELECT in("collect").name[0], in("collect").indexX[0], in("collect").indexY[0], acoustic, out("video").videoPath[0], out("video").videoDurationSec[0] FROM sensorrawdata WHERE acoustic>15 AND out("next").acoustic[0]>15 AND out("next").out("next").acoustic[0]>15 ORDER BY name
\end{lstlisting}
\hspace{5pt}
\textit{Neo4j Query:}
\begin{lstlisting}
MATCH (sn:sensornode)-[:collect]->(sa:sensorrawdata)-[:next]->(sb:sensorrawdata)-[:next]->(sc:sensorrawdata)-[:video]->(sv:sensorrawvideodata) WHERE sa.acoustic>15 AND sb.acoustic>15  AND sc.acoustic>15 RETURN sn.name, sa.acoustic, sn.indexX, sn.indexY,        sv.videoPath, sv.videoDurationSec ORDER BY sn.name
\end{lstlisting}
\hspace{5pt}
\textit{Relational SQL Query:}
\begin{lstlisting}
SELECT s.name, s.indexx, s.indexy, ra.collectDate, ra.acoustic, v.video_path, v.duration FROM sensornode s, sensorrawdata ra, sensorrawdata rb, sensorrawdata rc, sensorrawvideodata v WHERE ra.acoustic>15 AND rb.acoustic>15 AND rc.acoustic>15 AND s.id = ra.sensornode_id and ra.id = rb.next_id and rb.id = rc.next_id and rc.video_id = v.id ORDER BY      s.name ASC
\end{lstlisting}

\subsubsection{Recursive Query}
This query finds the detected “Human” typed objects with high accuracy and calculates the distance of the sensor node to the sink node.

\vspace{5pt}
\textit{OrientDB Query:}
\begin{lstlisting}
SELECT $nodeId, out('fusedby').name[0], fusionDate, $deep.count FROM fuseddata LET $nodeId = out('fusedby').@rid, $deep = SELECT COUNT(*) FROM (TRAVERSE in('lead') FROM $nodeId) WHERE concept = 'Human' AND weight > 0.9
\end{lstlisting}
\hspace{5pt}
\textit{Neo4j Query:}
\begin{lstlisting}
MATCH p=(a:sink)-[:lead*]->(b:gateway) WITH b.name as gname, length(p) AS depth MATCH  (sfd:fuseddata)-[:fusedby]->(sn:sensornode)<-[:lead]-(g:gateway) WHERE sfd.concept = "Human" AND  sfd.weight>0.9 AND g.name = gname RETURN gname, sn.name, sfd.fusionDate, depth
\end{lstlisting}
\hspace{5pt}
\textit{Relational SQL Query:}
\begin{lstlisting}
WITH RECURSIVE search_graph(id, name, lead, depth) AS SELECT g.id, g.name, g.lead, 0 FROM gateway g WHERE g.lead is null UNION ALL
\end{lstlisting}
\begin{lstlisting}
SELECT g.id, g.name, g.lead, 0 FROM gateway g WHERE g.lead is null UNION ALL
\end{lstlisting}
\begin{lstlisting}
SELECT g.id, g.name, g.lead, sg.depth + 1 FROM gateway g, search_graph sg WHERE g.lead = sg.id
\end{lstlisting}
\begin{lstlisting}
SELECT s.name, s2.name, s2.indexX, s2.indexY, s.depth FROM search_graph s inner join
\end{lstlisting}
\begin{lstlisting}
select distinct sn.id, sn.name, sn.indexX, sn.indexY, sn.lead FROM sinkfuseddata sfd, gatewayfuseddata gfd, sensorfuseddata srfd, sensornode sn WHERE srfd.fusion = gfd.id AND gfd.fusion = sfd.id AND srfd.fusedby = sn.id AND
sfd.concept = 'Human' AND sfd.weight>0.9
\end{lstlisting}
\begin{lstlisting}
s2 ON s2.lead = s.id
\end{lstlisting}

Table \ref{tab:4} shows the performance results of our example queries. For the first query is a simple range query which is focused on the basic query performance. OrientDB performs better than Neo4j because of the under-hood architecture of OrientDB which is a multi-model graph database. And graph model performs better than the relational model since there is no join operation as promised by the NoSQL databases.

For the second query, Neo4j performs better than OrientDB and the graph model is again better than MySQL. This time again graph databases beat relational databases because of their join-free query capability. To understand why Neo4j is faster than OrientDB, we have to dive into queries. The query is neighbors and neighbors of neighbors query, which is a typical graph matching problem considering paths of length 1 or 2. In PostgreSQL we use a relational table with id from / id to backed by an index. Neo4j performs better because of its “index-free adjacency” for the edges.

The last query is to test the recursive SQL type of a query. The graph-based model is much faster than the relational model. Neo4j fails for this recursive query. There can be some optimizations possible while using the Cypher language (the query language of Neo4j) but we couldn't find them in the available Neo4j documentation. 

\begin{table}
	\caption{Test Results (Numbers are in milliseconds)}
	\label{tab:4}
	\renewcommand{\tabcolsep}{0.7pc} 
	\renewcommand{\arraystretch}{1.1} 
	\begin{tabular}{llll}
		\hline
		Query&OrientDB&Neo4j&MySQL\\
		\hline
		Concept Based  & 209 &	618	& 938 \\
		Video Based	& 355 &	145	& 422	\\    
		Recursive	& 4,293 &	79,812 &	36,469 \\  
		\hline
	\end{tabular}
\end{table}

\subsection{Doubling Sensed Raw Data Size}
An OrientDB graph database is selected for the execution of experiments. The previous experiments are applied on generated synthetic data of one month where each sensor node can sense data with 5 minutes of period. Now we increase the simulation duration from 1 month to 5 months step by step and diagnosed the query performance.

\begin{figure}
	\includegraphics[scale=0.96]{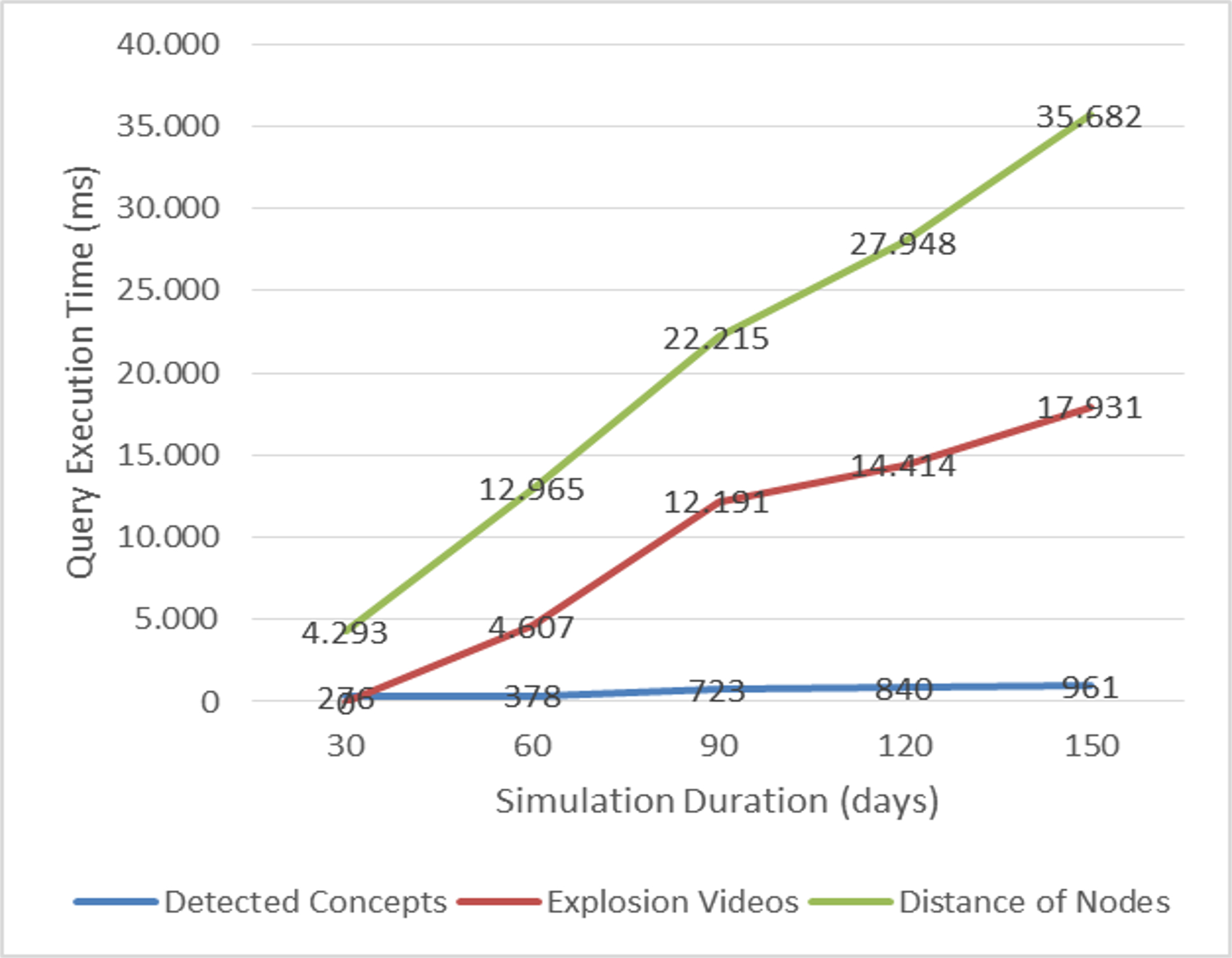}
	\caption{OrientDB Query Time/Simulation Duration Chart}
	\label{fig:8}
\end{figure}

Figure \ref{fig:8} shows the chart of query times affected by the increased simulation duration. The query time is increased and it is better than linear which is fairly well compared to the doubled data size. 

\section{Conclusions}
\label{sec:7}
In this paper, we propose a graph-based model for complex multimedia and sensor big data.
Our first aim is to represent the wireless sensor networks
with multimedia data. Our implementation is
composed of two main modules. The first module is
the implementation of the proposed data model. The second
module is the simulation which includes a simulator
to produce synthetic big sensor multimedia data and the
simulation infrastructure which represents the objects
moving in our multimedia sensor networks.

We have focused on the surveillance systems to design
our graph data model. The network topology and
the data flow between each sensor nodes, gateways
and sink are modeled so that all static and kinetic data
is stored within the sensor network which must be assumed
to be big data. The database to store big
data is the NoSQL graph database. Because graph databases are
good at representation of complex relations and scalable
to store big multimedia sensor data.

We have tested our model on both graph databases and a relational database. Test results show that graph database model performs better that the relational database model. To decide which graph database is more convenient and efficient, we chose two well-known graph databases, OrientDB and Neo4j, for experiments. Neo4j is the market leading graph database, but it does not fit into our needs, which is to store complex multimedia big data. Because Neo4j supports high availability with the master-slave approach, which can scale vertically. But, in order to survive in the big data world, we need a master-master approach, which OrientDB has. In addition to that, our
experiments show that OrientDB performs better than Neo4j.

We have simulated our WMSN prototype system with millions of data to test the proposed graph model simulating the system. We have tested the query performance with many complex scenarios. We have showed that our generated millions of synthetic data can be efficiently queried efficiently on our graph database.

As future research work, we plan to do graph analytics on our big graph model. Object tracking, topology optimization, and path extraction like analytics suit well for our surveillance domain.

\section*{Acknowledgements}
This work is supported in part by a research Grant from TÜBİTAK with Grant No. 114R082. We thank to the researchers of CEng Multimedia Database Lab. for their very valuable comments.
\section*{References}

\bibliography{referenceBibtex}

\end{document}